\documentclass[a4paper]{jpconf}
\usepackage{graphicx,epsf} %,amssymb,amsmath,latexsym}
\newcommand{\cu}{\mbox{${\mathcal U}$}}
\newcommand{\be}{\begin{eqnarray}}
\newcommand{\ee}{\end{eqnarray}}

\renewcommand{\d}{\mbox{{\rm d}}}

\def \cu{{\cal U}}

\begin{document}
\title{Gravitational collapse and evolution of holographic black holes}
\author{R Casadio$^1$, C Germani$^2$}
\address{$^1$Dipartimento di Fisica, Universit\`a di Bologna and I.N.F.N.,
Sezione di Bologna, via Irnerio~46, 40126 Bologna, Italy}
\address{$^2$D.A.M.T.P., Centre for Mathematical Sciences,
University of Cambridge, Wilberforce road, Cambridge CB3 0WA, England}
\ead{casadio@bo.infn.it,C.Germani@damtp.cam.ac.uk}
\begin{abstract}
Gravitational collapse is analyzed in the Brane-World by arguing
that regularity of five-dimensional geodesics require that stars
on the brane have an atmosphere.
For the simple case of a spherically symmetric cloud
of non-dissipating dust, conditions are found for which the
collapsing star evaporates and approaches
the Hawking behavior as the (apparent) horizon is being formed.
The effective energy of the star vanishes at a finite radius
and the star afterwards re-expands and ``anti-evaporates''.
Israel junction conditions across the brane
(holographically related to the matter trace anomaly) and the
projection of the Weyl tensor on the brane (holographically
interpreted as the quantum back-reaction on the brane metric)
contribute to the total energy as, respectively, an ``anti-evaporation''
and an ``evaporation'' term.
\end{abstract}
\section{Introduction}
\label{intro}
Black holes (BHs) are unstable in four (and higher) dimensions because
of the Hawking effect~\cite{hawking}, which is deeply linked
to the trace anomaly of radiation fields~\cite{birrell}.
In the Randall-Sundrum (RS) Brane-World (BW)
models~\cite{RS}, a collapsing homogeneous star likewise requires
a non-static exterior~\cite{BGM}.
Further, forcing a static exterior induces a trace anomaly of the same
form as that of semiclassical BHs, although with opposite sign, which
suggested that BH solutions of the bulk equations correspond
to quantum corrected (semiclassical) BHs on the brane~\cite{tanaka1,fabbri},
in the spirit of the holographic principle~\cite{holography} and AdS/CFT
conjecture~\cite{AdSCFT}.
\par
\begin{figure}[ht]
\begin{minipage}{18pc}
\includegraphics[width=15pc]{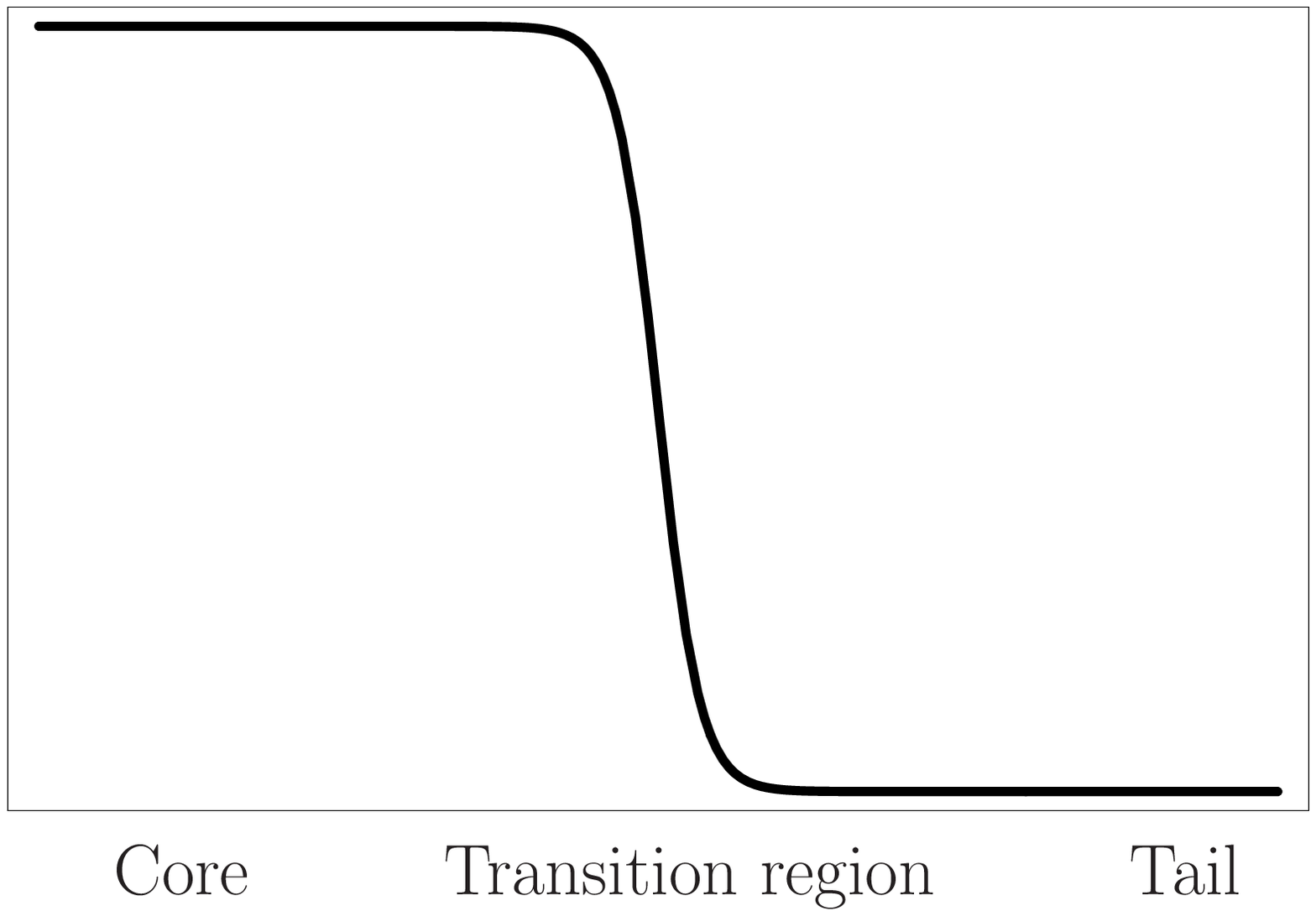}
\caption{Density profile.
\label{figura}}
\end{minipage}
\hspace{1pc}
\begin{minipage}{18pc}
\raisebox{3.5cm}{$V$}
\includegraphics[width=15pc]{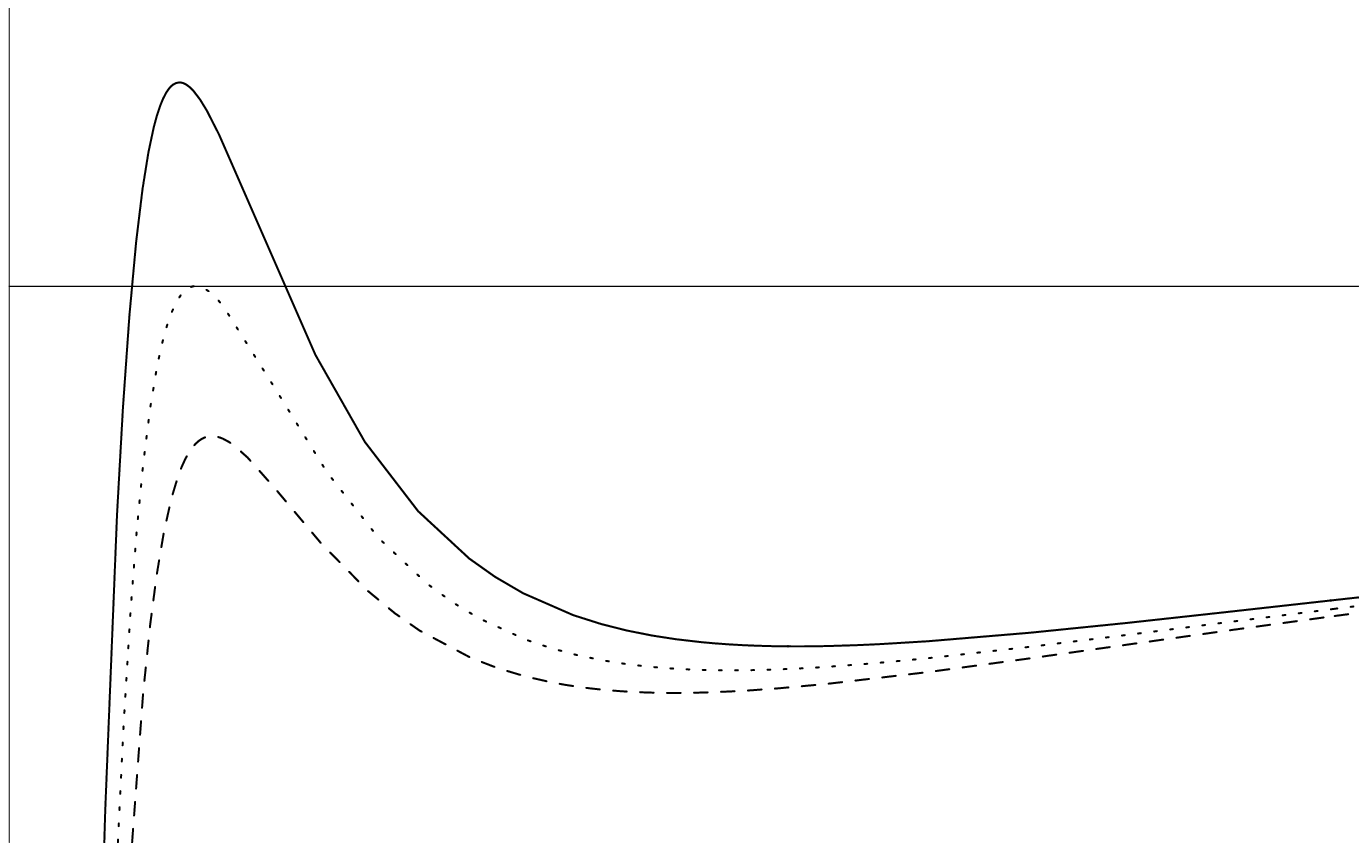}
{\raisebox{0cm}{$X$}}
\caption{Qualitative behavior of $V$ for $\mu>\mu_{\rm c}$ (solid line)
and $\mu<\mu_{\rm c}$ (dashed line).
For $\mu=\mu_{\rm c}$ (dotted line), the peak of $V$ equals the
shells energy $E=0$.
\label{V(X)}}
\end{minipage} 
\end{figure}

The junction conditions~\cite{israel}, which preserve
the regularity of (five-dimensional) geodesics, cannot allow a
step-like discontinuity (e.g.~across the star surface) and
a $\delta$-like discontinuity (e.g.~across the brane)
in the stress tensor at the same location,
hence discontinuities in the stress tensor of brane stars are not
mathematically permitted.
This can be physically understood by considering that the brane
thickness (of the order of the AdS length $\ell\sim\lambda^{-1/2}$,
$\lambda$ being the brane tension)
and the star's atmosphere cannot be both negligibly thin,
and it is indeed natural to assume that the latter is
much larger than $\ell$.
In Ref.~\cite{PTP}, we employed effective four-dimensional
(hydrodynamical) equations~\cite{maart} in order to study a
``corrected'' Oppenheimer-Snyder (OS) model~\cite{OS}
in which the star is divided into three regions
(see Fig.~\ref{figura}):
a homogeneous ``core'' with most of the energy;
a ``transition region'' of fast density decrease which connects
with a ``tail'', where the energy density slowly vanishes.
The tail and transition region together form the ``BW atmosphere'',
which disappears in the limit $\lambda\to\infty$.
\par
In general, effective equations cannot determine the brane
metric uniquely unless one also knows the bulk geometry.
However, for negligible dissipation
and asymptotically flat brane, the evolution of the system
is uniquely determined by the dynamics of the homogenous core, which
corresponds to an exact five-dimensional solution~\cite{langlois}
and reproduces the trace anomaly of quantum field theory~\cite{birrell}.
Our main results are that {\em the total energy of the
system is conserved\/} and that {\em the collapsing star ``evaporates''
until the core experiences a ``rebound'' in the high energy regime
(when its energy density is comparable with $\lambda$), after
which the whole system ``anti-evaporates''\/}.
This behaviour cannot be related to GR perturbatively
(in $\epsilon\equiv {\rho_0/ \lambda}$, where
$\rho_0$ is the initial core density),
but it seems in agreement with the uncertainty principle of
quantum mechanics~\cite{impBO}.
Moreover, we can find a range of parameters for which the
minimum radius of the collapsing core is larger than $\ell$
(which sets the scale of Quantum Gravity in the BW),
thus supporting the holographic interpretation of the model.
\section{Spherically symmetric collapsing dust}
\label{coll}
Effective four-dimensional Einstein equations for dust
and vanishing brane cosmological constant can be written
as~\cite{maart}
$G_{\mu\nu}=8\,\pi\,T_{\mu\nu}^{\rm eff}=
8\,\pi\,\left(\rho^{\rm eff}\,u_\mu\,u_\nu
+p^{\rm eff}\,h_{\mu\nu}
+\Pi_{\mu\nu}\right)$,
where $u^\mu$ is the unit four-velocity of dust, $h_{\mu\nu}$
the space-like metric that projects orthogonally to $u^\mu$,
\be
\rho^{\rm eff}=\rho\,\left(1+\frac{\rho}{2\,\lambda}\right)
+{\cal U}
\ ,
\quad
p^{\rm eff }=\frac{\rho^2}{2\,\lambda}+\frac{\cal U}{3}
\ ,
\ee
with $\rho$ the (``bare'') energy density, and $\cu$ and
$\Pi_{\mu\nu}$ come from the Weyl tensor projected on the
brane~\footnote{We set the momentum density $Q_\mu=0$,
see Ref.~\cite{PTP} for more details.}.
Bianchi identities supplied by the junction conditions then
produce both local (LCE) and non-local conservation equations
(NLCE).
For negligible dissipation, one can take a Tolman brane geometry,
$\d s^2=-\d\tau^2+\left(R'\right)^2\,\d r^2+R^2\,\d\Omega^2$
where $R=R(\tau,r)$, and the LCEs imply conservation of the
``bare'' mass function
\be
m_\rho(r)\equiv \frac{4\,\pi}{3}\,\int^{r}_0
\rho(\tau,x)\,\partial_x\left(R^3(\tau,x)\right)\,\d x
\ .
\label{M_0}
\ee
The system of NLCEs is in general not closed, since we do not have
an equation for $\Pi_{\mu\nu}$.
However, for sufficiently large $R$, the knowledge of $\Pi_{\mu\nu}$
in an extended spatial region together with the asymptotic
flatness and the continuity of the Weyl tensor make that
system closed.
Another important result which follows from the LCEs and NLCEs is
that, {\em if the brane metric is asymptotically flat, the anisotropic
stress $\Pi_{\mu\nu}\not=0$ whenever $\dot\rho'\neq 0$\/}.
\par
The $(\tau,\tau)$ Einstein equation yields the equation of motion
\be
\dot R^2(\tau,r)=\frac{2\,M(\tau,r)}{R(\tau,r)}
\ .
\label{R}
\ee
where the (in general time-dependent) ``effective'' mass
\be
M(\tau,r)=\frac{4\,\pi}{3}\,\int^{r}_0
\rho^{\rm eff}(\tau,x)\,\partial_x \left(R^3(\tau,x)\right)\,
\d x
\ ,
\label{Meffg}
\ee
replaces the bare mass of General Relativity (GR).
The latter is always well defined and dust shells in GR move along geodesics
to reach the central singularity ($R=0$) at increasing proper times
(Tolman model~\cite{tolman}) or at the same proper time (OS model~\cite{OS}).
In the BW, the effective mass $M$ drives the collapse, but it diverges
for $R\to 0$ and this would make the whole four-dimensional brane singular.
In order to avoid this (physically unlikely) case, one has to include
a sufficiently negative contribution to the mass from the projected Weyl
tensor which will generate an Hawking flux near the forming horizon and
make the effective mass evaporate completely.
\subsection{Core}
\label{s_core}
The bulk solution which corresponds to the OS
core of the star is regular in five dimensions~\cite{langlois}
and, since $\rho'= 0$, the system of relevant equations is now
closed with $\Pi=0$ and
\be
\cu=-\frac{27\,\mu\,r^4\,\epsilon}{128\,\pi^2\,r_0^4\,\rho_0\,R^4}
\ ,
\label{Ocu}
\ee
where $\mu$ is a constant.
The physical radius $R$ can be written in the factorized form
\be
R(\tau,r)=\left(\frac{9}{2}\,M_{\rm S}\right)^{1/3}\frac{r}{r_0}
\,X(\tau)
\ ,
\label{g(r)}
\ee
in which $M_{\rm S}$ is the total bare mass of the OS core,
and the effective mass~(\ref{Meffg}) is given by
\be
M(\tau,r)=M_{\rm S}\left(\frac{r}{r_0}\right)^3
+\frac{9\,\epsilon}{32\,\pi\,\rho_0}
\left(\frac{r}{r_0}\right)^4\left[
\frac{(2\,M_{\rm S})^2}{3\,R^3}\left(\frac{r}{r_0}\right)^2
-\frac{\mu}{R}\right]
\ ,
\label{Mos}
\ee
where the second term represents a BW correction.
The above mass would diverge for $R\to 0$, the harmless
(at least when covered by an horizon) central singularity of
GR, and make the whole space-time singular (this also occurs
for a more general Tolman core~\cite{PTP}).
One must therefore have $\mu$ positive and large enough so that each
shell will bounce back after reaching a minimum radius $R_{\rm min}$
where the corresponding effective mass vanishes.
The Weyl tensor, holographically interpreted as the quantum
back-reaction on the brane metric, contributes the ``evaporation''
term proportional to $\mu$ which dominates at low energies;
the BW correction to the matter stress tensor, holographically
interpreted as the matter quantum trace anomaly~\cite{tanaka1},
yields the ``anti-evaporating'' term proportional to $M_S^2$
which increases with the energy.
\par
\begin{figure}[t]
\begin{minipage}{18pc}
\includegraphics[width=15pc]{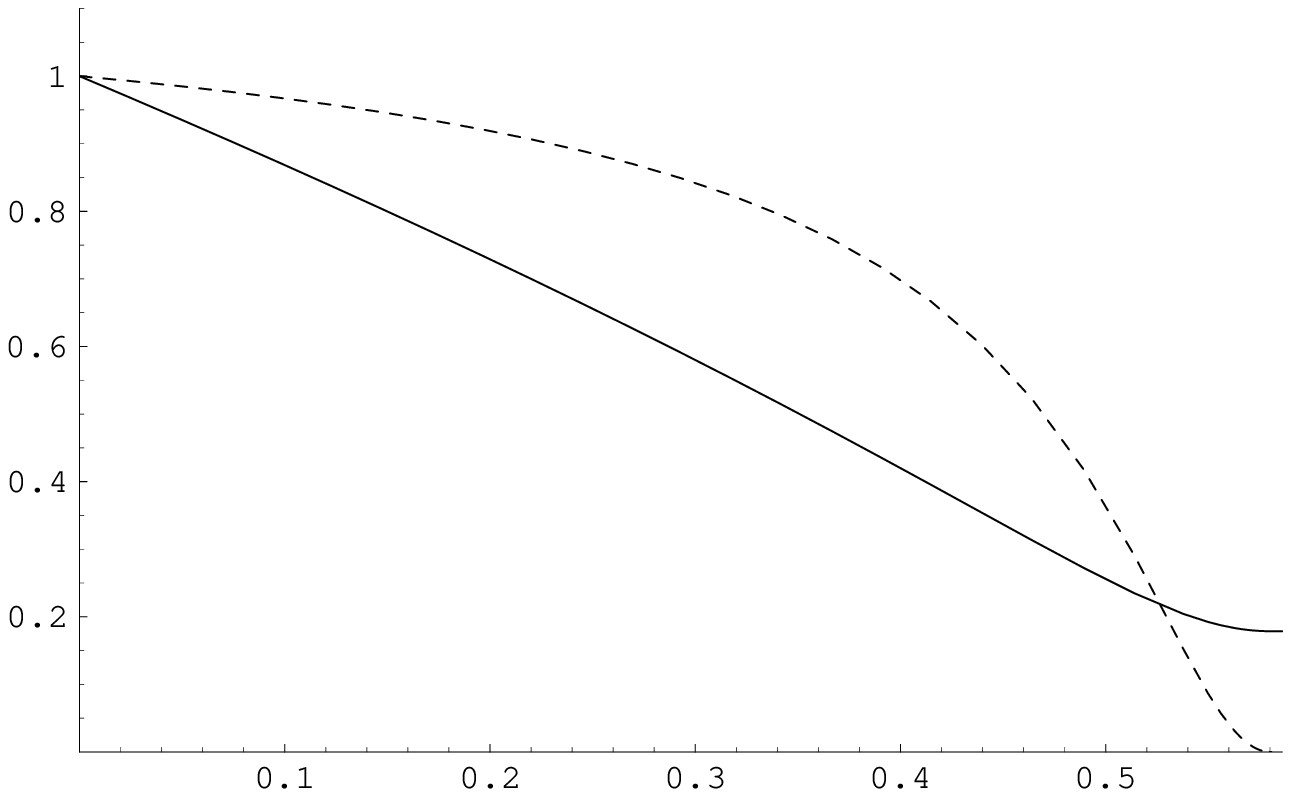}
\caption{Typical evolutions of the core radius $R_0(\tau)/R_0(0)$
(solid line) and effective mass $M_0(\tau)/M_0(0)$ (dashed line) for
$\mu>\mu_{\rm c}$.
Time is arbitrary.
\label{mu<mu_c}}
\end{minipage}
\hspace{1pc}
\begin{minipage}{18pc}
\raisebox{3.5cm}{$\frac{\d M_{\rm H}}{\d t}$}
\includegraphics[width=15pc]{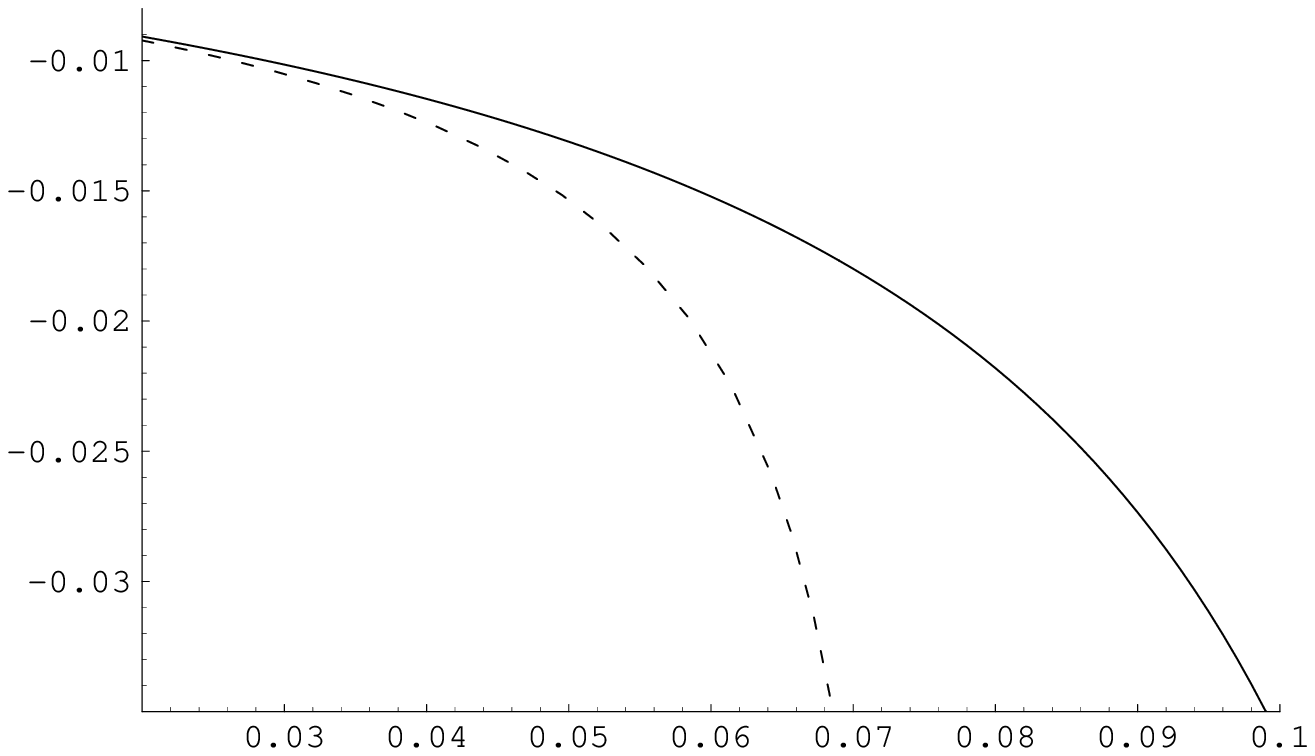}
{\raisebox{0cm}{$t$}}
\caption{Energy flux versus the time $t=(\tau-\tau_{\rm H}^{\rm OS})/T$
(solid line) compared to the Hawking flux (dotted line)
for $\epsilon=10^{-4}$, $M_{\rm S}=\rho_0=1$, $T=10$ and
$\mu=5000>4824=\mu_{\rm c}$.
\label{dMH0}}
\end{minipage} 
\end{figure}

Upon inserting the effective mass~(\ref{Mos}) into Eq.~(\ref{R}),
one obtains
\be
\dot X^2=
\frac{4}{9\,X}
+\frac{\epsilon}{27\,\pi\,\rho_0\,X^{4}}
-\frac{6^{1/3}\,\epsilon\,\mu}{24\,\pi\,\rho_0\,M_{\rm S}^{4/3}\,X^{2}}
\equiv	
-V(X)
\ ,
\label{Xeq}
\ee
which shows that the core remains ``rigid'' through the bounce
and it is thus sufficient to consider the evolution of its surface
at $r=r_0$.
There exists a critical value
$\mu_{\rm c}=\left(32\,\pi\,\rho_0\,M_{\rm S}^2/{3\,\epsilon}\right)^{2/3}$
such that the bounce occurs only for $\mu>\mu_{\rm c}$ (for $\mu=\mu_{\rm c}$
the two turning points coincide, see Fig.~\ref{V(X)}).
In Fig.~\ref{mu<mu_c} we display a typical trajectory of
$R_0(\tau)=R(\tau,r_0)$, along with the corresponding
effective mass $M_0(\tau)=M(\tau,r_0)$, for $\mu>\mu_{\rm c}$.
After the rebound, the system reverses its evolution,
however, in a more realistic model, the collisionless description
of dust should be relaxed and dissipation from both the core
and the atmosphere is expected to make the process irreversible.
\par
Since $\dot R_0<0$ and $R_0(0)\gg 2\,M_{\rm S}$ (there is no initial
horizon), for $\mu>\mu_{\rm c}$
\be
\dot M_0(\tau)=
-\frac{9\,\epsilon}{32\,\pi\,\rho_0}\,
\left[\left(\frac{2\,M_{\rm S}}{R_0}\right)^2
-\mu\right]\,\frac{\dot R_0}{R_0^2}
<0
\ ,
\label{dM0}
\ee
and the evaporation sets out at the beginning of the collapse
and goes on at least until the radius bounces back at
$R_{\rm min}\sim
\ell\,\left(
M_{\rm S}/\ell
\right)^{1/3}
\gg
\ell
\sim
\lambda^{-1/2}$.
Since $R_{\rm min}$ is the shortest length in our system,
the holographic description is expected to hold for
astrophysical sources for which
$M_{\rm S}\gg\lambda^{-1/2}$
(it also follows that $\mu_c\gg 1$).
\par
The horizon trajectory $r=r_{\rm H}(\tau)$ is defined as the value
of $r$ at which the horizon is formed at the time $\tau$, that is
$R(\tau,r_{\rm H}(\tau))=2\,M(\tau,r_{\rm H}(\tau))$.
The time derivative of the horizon effective mass,
$M_{\rm H}(\tau)\equiv M(\tau,r_{\rm H}(\tau))$,
\be
\frac{\d M_{\rm H}}{\d\tau}=\dot M_{\rm H}+M_{\rm H}'\,\dot r_{\rm H}
\ ,
\label{dMHtot}
\ee
then contains two contributions: one, from the intrinsic time
dependence of $M$ at constant $r$, is a first order effect in
$\epsilon$ which vanishes in GR;
the second one, due to the (possibly) variable number of shells
included within the horizon, depends on the detailed form of the
atmosphere.
\par
Since in the core, the velocity $|\dot R(\tau,r)|$ increases monotonically
in $r$ at fixed $\tau$, an (apparent) horizon forms at the boundary
$r=r_0$ at $\tau=\tau_{\rm H}^{\rm OS}\equiv T-\frac{4}{3}\,M_{\rm S}$,
where $T$ fixes the time scale of the collapse.
From Eq.~(\ref{dM0}), we then get the (instantaneous) Hawking flux~\cite{hawking}
\be
\frac{\d M_{\rm H}}{\d\tau}
\simeq
-\frac{9\,\left(\mu-1\right)\,\epsilon}{128\,\pi\,\rho_0\,M_{\rm H}^2}
\ ,
\label{MRI}
\ee
precisely at the time $\tau=\tau_{\rm H}^{\rm OS}$.
\subsection{Transition region}
In the GR model $\rho=0$ for $r>r_0$.
The density must therefore decrease rapidly from $\rho=O(\epsilon^0)$
at $r=r_0$ to $\rho=O(\epsilon)$ for $r=r_{\rm s}$.
Moreover, since the transition is a BW effect, we can take
$r_{\rm s}-r_0=O(\epsilon)$ as well as $\cu=O(\epsilon)$.
Combining these results, we obtain, to first order in $\epsilon$,
\be
\dot M(\tau,r)\simeq \dot M_0(\tau)
\ ,
\label{mII}
\ee
for $r_0<r<r_{\rm s}$, and the Hawking flux will remain
negative (and substantially unaffected) throughout the border
of the transition region $r=r_{\rm s}$.
\subsection{Tail}
As in the transition region, $\rho=O(\epsilon)$ for $r_{\rm s}<r$,
and $\rho'\,\rho/\lambda=O\left(\epsilon^2\right)$,
so that bulk gravitons are decoupled from brane matter.
The Weyl contribution is however of the same order,
$m_\cu(\tau,r;r_{\rm s})=
\frac{4\,\pi}{3}\,\int_{r_{\rm s}}^r
\cu\,\left(R^3\right)'\,\d x
=O(\epsilon)$,
and satisfies
\be
\dot{m}_\cu+2\,\sqrt{\frac{2\,M_{\rm S}}{R^3}}\,m_\cu=
\dot M_0\,\left[\left(\frac{R_0}{R}\right)^{3/2}-1\right]
\ .
\label{dmeq}
\ee
Since $R(\tau,r)>R_0(\tau)$ for $r>r_0$, (\ref{dmeq}) implies that
$m_\cu$ cannot remain zero in the tail.
\par
In order to proceed, we now assume that:
\begin{description}
\item[(i) ]
{\em $\lim_{r\to\infty} R(\tau,r)=\infty$}, and
\item[(ii)]
{\em the effective mass be finite at spatial infinity\/},
$\lim_{r\to\infty} M(\tau,r)<\infty$,
$\forall\,\tau>0$.
\end{description}
Since $M_0(\tau)$ always remains finite if $\mu>\mu_{\rm c}$
and the tail bare mass is small by construction,
this implies that
$\lim_{r\to\infty} m_\cu(\tau,r;r_{\rm s})<\infty$,
$\forall\,\tau>0$.
Asymptotic flatness ensures that we can take the limit
$r\to\infty$ (equivalent to $R\to\infty$ at fixed time)
in Eq.~(\ref{dmeq}) and finally obtain
\be
\lim_{r\to\infty} \dot{m}_\cu(\tau,r;r_{\rm s})+\dot M_0(\tau)
=\lim_{r\to\infty} \dot M(\tau,r)=0
\ ,
\ \ \ \forall\,\tau>0
\ .
\label{dMinfty}
\ee
We have thus shown that {\em if the total effective mass at spatial
infinity is finite at the initial time $\tau=0$, it will always remain
constant\/} (for a bouncing core evolution with $\mu>\mu_{\rm c}$), so that
{\em the total effective mass of the collapsing dust star is actually
conserved\/}.
\par
It is particularly interesting to consider the case for which there is
initially no energy stored in $\cu$.
We then see from Eq.~(\ref{dmeq}) that $m_\cu>0$ during the collapse
and, after the bounce, we expect that $m_\cu$ will also evolve backwards
so as to ensure the condition (\ref{dMinfty}).
The energy flux along the horizon trajectory $r=r_{\rm H}(\tau)$
can also be obtained numerically for this case and is plotted in
Fig.~\ref{dMH0} for some values of the parameters.
We can see that it grows less with respect to that predicted by Hawking
and the conclusion is that, although the evaporation sets out
according to Hawking's law, the back-reaction on the brane
metric subsequently reduces the emission.
\section{Luminosity}
A distant observer experiences an impinging flux of energy
during the collapse, 
\be
\Phi_t\equiv
\frac{\d}{\d\tau}\left[\lim_{\bar r\to\infty}
\frac{4\,\pi}{3}\,\int^{\bar r}_{r_{\rm H}(\tau)}
\rho^{\rm eff}\,\left(R^3\right)'\,\d r\right]
\simeq
-\frac{\d M_{\rm H}}{\d\tau}
\ ,
\ee
which shows the same dependence on the mass $M_{\rm H}$ as the
semiclassical expression when the horizon is first forming, and
subsequently decreases to zero (before it becomes negative).
However, since this happens after the apparent horizon begins to
form, a distant observer might have to wait an infinite amount
of time to measure a vanishing flux.
\par
Upon equating the BW result (\ref{MRI}) at the time when the horizon
starts to form to the semiclassical luminosity calculated in the
Schwarzschild background for an astrophysical object, we obtain a
bound for the AdS length~\cite{PTP},
\be
10^{-32}\, {\rm mm}\ll \ell \ll 10^{-9}\, {\rm mm}
\ ,
\label{sb}
\ee
which is three orders of magnitude better than the best constraint found
in~\cite{emparan} considering the time scale of primordial BH evaporation.
\section{Trace anomaly}
Strictly speaking, there is no trace anomaly in our approach,
since we have included the back-reaction of the effective matter
on the brane metric.
However, in order to compare with known results {\em without\/}
the back-reaction, we can define the trace anomaly ${\mathcal R}$
as the sum of the Ricci scalar and the trace of the {\em bare\/}
stress tensor.
At the OS boundary, $r=r_0$, we then have
\be
{\mathcal R}=
-\frac{9}{2\,\pi\,\lambda}\,\frac{M_{\rm S}^2}{R^6}
\ ,
\ee
which is the quantum Ricci anomaly of Ref.~\cite{birrell} with
the correct sign.
It is then clear that the sign mismatch found in Ref.~\cite{BGM}
was due to the choice of a non-smooth energy density and that
the atmosphere describes how the excess energy stored
in the OS boundary is released.
\section{Conclusions}
\label{conc}
Inspired by the conjecture that classical BHs in the BW may
reproduce the semiclassical behavior of four-dimensional BHs,
we have studied the gravitational collapse of a spherical star
of dust in the RS scenario.
Regularity of the bulk geometry requires continuity of the matter
stress tensor on the brane and leads to a loss of mass from
the boundary of the star.
We found that the system of effective BW equations is closed to
our level of approximation and leads to the collapsing dust star
emitting a flux of energy which, at relatively low energies,
approaches the Hawking behavior when the (apparent) horizon is being
formed.
However, no real space-time singularity forms since the star effective
mass vanishes at finite star radius, thus leaving a remnant which
re-expands by absorbing back the emitted radiation.
In a more realistic case, one however expects that dissipation cannot
be neglected and the process then becomes irreversible.
\par
Let us finally point out that all the above features were obtained
for BHs formed by gravitational collapse, excluding therefore
primordial BHs.
\section*{References}
\end{document}